\documentclass[fleqn,10pt]{ScienceAdvances}
\definecolor{dkgreen}{rgb}{0,0.6,0}

\usepackage{graphicx}
\usepackage{amsmath,bm}
\usepackage{comment}
\usepackage{mathtools}
\usepackage{cellspace}
\usepackage{multirow}
\usepackage{caption}
\usepackage{makecell}
\usepackage{sectsty}
\usepackage{hyperref}

\title{Homophily and minority size explain perception biases in social networks}

\author[1,+,*]{Eun Lee}
\author[2,3,+,*]{Fariba Karimi}
\author[2,3]{Claudia Wagner}
\author[4,5,6]{Hang-Hyun Jo}
\author[2,7]{Markus Strohmaier}
\author[8,9]{Mirta Galesic}

\affil[1]{Department of Mathematics, University of North Carolina at Chapel Hill, 27599, USA}
\affil[2]{Department of Computational Social Science, GESIS, Cologne, 50667, Germany}
\affil[3]{Institute for Web Science and Technologies, University of Koblenz-Landau, Koblenz, Germany}
\affil[4]{Asia Pacific Center for Theoretical Physics, Pohang 37673, Republic of Korea}
\affil[5]{Department of Physics, Pohang University of Science and Technology, Pohang 37673, Republic of Korea}
\affil[6]{Department of Computer Science, Aalto University, Espoo FI-00076, Finland}
\affil[7]{HumTec Institute, RWTH Aachen University, Germany}
\affil[8]{Santa Fe Institute, Santa Fe, NM}
\affil[9]{Harding Center for Risk Literacy, Max Planck Institute for Human Development, Berlin, Germany}

\affil[+]{These authors contributed equally to this work.}
\affil[*]{Corresponding authors. Email: fariba.karimi@gesis.org, eunfeel@email.unc.edu}


\begin{document}



\captionsetup[figure]{labelfont={bf},name={Fig.},labelsep=period} 

\flushbottom
\maketitle
\section*{Abstract} 

People's perceptions about the frequency of attributes in their social networks sometimes show \emph{false consensus}, or overestimation of the frequency of own attributes, and sometimes \emph{false uniqueness}, or underestimation of the frequency.  Here we show that both perception biases can emerge solely from the structural properties of social networks. Using a generative network model, we show analytically that perception biases depend on the level of homophily and its asymmetric nature, as well as on the size of minority group. Model predictions correspond to empirical data from a cross-cultural survey study and to numerical calculations on six real-world networks. We also show in what circumstances individuals can reduce their biases by relying on perceptions of their neighbors. This study advances our understanding of the impact of network structure on perception biases and offers a quantitative approach for addressing related issues in society. 

%
%
\thispagestyle{empty}


\section*{Introduction}

People's perceptions of their social worlds determine their own personal aspirations\cite{Cialdini1998} and willingness to engage in different behaviors, ranging from voting\cite{Bond2012} and energy conservation\cite{Allcott2011} to health behavior\cite{Centola2010}, drinking\cite{Borsari2003} and smoking\cite{Botvin1992}.
Yet when forming these perceptions, people seldom have an opportunity to draw representative samples from the overall social network. Instead, their samples are constrained by the local structure of their personal networks, which can bias their perception of the frequency of different attributes in the general population. For example, during the 2016 U.S. presidential election Twitter users, including journalists, who supported one candidate isolated themselves from users who supported a different candidate\cite{isolated_journalists}. Members of such insular communities often tend to overestimate the frequency of their own attributes in the overall society. Such \emph{social perception biases} have been studied in the literatures on false consensus\cite{Ross1977,Mullen1985,Krueger1994,Krueger2007}.
However, it has also been documented that people holding a particular view sometimes tend to underestimate the frequency of that view. This apparently contradictory bias has been studied in the literature on false uniqueness\cite{frable1993being,Mullen1992,Suls1987}, pluralistic ignorance\cite{Miller1987,Prentice1993} and majority illusion\cite{lerman2016majority}. 

It has been observed that diverse social perception biases can be related to the structural properties of personal networks\cite{Sherman1983, galesic2018sampling}, which can strongly affect the samples of information that individuals rely on when forming their social perceptions\cite{Juslin2007,pachur2013intuitive}. However, the impact of different network properties on social perception biases has not yet been systematically explored. Most existing explanations of these biases instead rely on assumptions about motivational and cognitive processes, including feeling good when others share one's own view\cite{marks1987ten}, wishful thinking\cite{miller2012citizen}, easier recall of the reasons for having own view\cite{Ross1977}, justifying one's undesirable behaviors by overestimating their frequency in the society\cite{Suls1988}, and rational inference of population frequencies based on own attributes\cite{dawes1989statistical}. These accounts can explain biases such as false consensus but not false uniqueness. 

Here we show empirically, analytically, and numerically how a simple network model can explain social perception biases resembling both
false consensus and false uniqueness, without further assumptions about biased motivational or cognitive processes. Empirical results from a cross-cultural survey show that structural network properties, namely, level of homophily and minority-group size, influence people's social perception biases. Analytical results from a generative network model with tunable homophily and minority-group size align well with the empirical findings. Numerical investigations show that model predictions are consistent with sample biases that could occur in six empirical networks, and point to the importance of an additional structural property, asymmetric homophily. We also show when perception biases can be reduced by aggregating one's own perceptions with the perceptions of one's neighbors. We discuss the implications of these results for the understanding of the nature of human social cognition and related social phenomena.

\section*{Results}
\subsection*{Defining social perception biases} 

We focus on perceptions of the prevalence of binary attributes (e.g., smoking, believing in a god, or donating to charity) in the overall network. We define social perception biases on the individual and the group level. Unless otherwise stated, we use $P_{\textrm{indv.}}$ and $P_{\textrm{group}}$ to denote individual- and group-level perception bias. If necessary, we add a superscript $\textrm{a}$ for the minority and $\textrm{b}$ for the majority group. 

On the individual level, we assume that individuals' perceptions are based on the prevalence of an attribute in their personal network (direct neighborhood). We define individual $i$'s bias in perception of the fraction of individuals with a given attribute in the overall network as follows:

\begin{equation}
P_{\textrm{indv.}} = \frac{1}{f_a} \frac{\sum_{j\in\Lambda_i}x_j}{k_i},
\label{eq:p_indiv}
\end{equation}

where $k_i$ is the degree of individual $i$, that is, the size of $i$'s personal network,  $\Lambda_i$ is a set of $i$'s neighbors, $x_j$ is the attribute of individual $i$'s neighbor $j$ ($x_j=1$ for a minority attribute and $0$ for a majority attribute), and $f_{a}$ is the true fraction of minority nodes in the overall network. 

The group-level perception bias is defined by aggregating perception biases of all individuals in a minority or a majority group:

\begin{equation}
P_{\textrm{group}} = \frac{1}{f_a} \frac{ \sum_{i\in N_g}\sum_{j\in\Lambda_i}x_j}{\sum_{i\in N_g} k_i},
\label{eq:p_group}
\end{equation}

where $N_g$ is the set of individuals in a group $g$, which can be a minority or a majority group. This definition is comparable to those used in studies of the friendship paradox~\cite{eomhang,hangpre} and enables us to derive analytic predictions at the group level using a mean-field approach. 

The minimum value of the perception bias is $0$ and the maximum value is $1/f_{a}$ (see Method). A perception-bias value below $1$ indicates an underestimation of the minority-group size, and a value above $1$ indicates an overestimation. If the value equals $1$ a group or an individual perfectly perceives the fraction of a minority attribute in the overall network. 

\begin{figure*}[ht]
\centering
\includegraphics[width=0.8\linewidth]{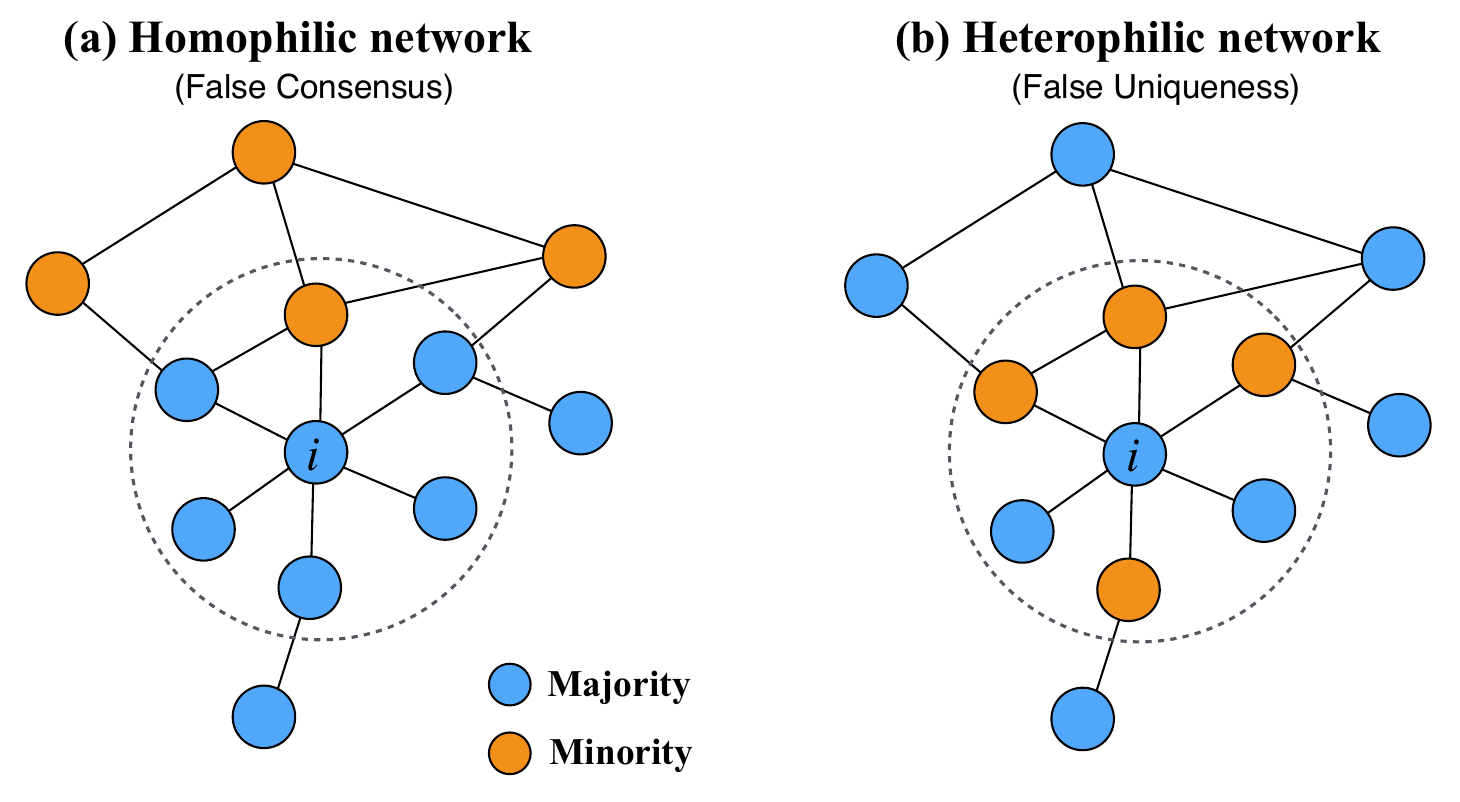}

\caption{\textbf{Individual- and group-level social perception bias.} Individuals belong to one of two groups: the majority (blue) or the minority (orange). The minority fraction is $1/3$ in both networks ($f_{a} \approx 0.33$). Panel (a) depicts a homophilic network and panel (b) shows a heterophilic network. We can study social perception biases originating on the individual and the group level.
On the individual level, individual $i$ perceives the size of the minority group in the overall network based on his personal network, denoted by dashed circles. 
In the homophilic network, $i$ perceives the size of the minority to be approximately $1/6 \approx 16\%$, while in the heterophilic network $i$ perceives the size of the minority to be approximately $4/6 \approx 67\%$. Therefore, in the homophilic network, individual $i$ underestimates the minority-group size by a factor of $0.5$ and in the heterophilic network $i$ overestimates the minority size by a factor of $2$.
On the group level, the majority group perceives the size of the minority group to be $4/18 \approx 20\%$ in the homophilic network and $10/18 \approx 56\%$ in the heterophilic network.
The majority group underestimates the size of the minority group by a factor of $0.6$ in the homophilic network and overestimates the minority-group size by a factor of $1.67$ in the heterophilic network. In sum, depending on the topological structure of the network, individuals' and groups' perceptions about their own and other groups' sizes can be distorted.}
\label{fig:model}
\end{figure*}

As an example, Fig.~\ref{fig:model} illustrates how we define the perception bias at the individual level and group level for a high-homophily (homophilic) network and a low-homophily (heterophilic) network. Here, the color of an individual node depicts its group membership. Orange nodes belong to the minority and blue nodes to the majority. We focus on the central individual $i$ who is in the majority in both networks. This individual estimates the size of the minority group on the basis of the fraction of orange nodes in his personal network (dashed circle). In the homophilic network (Fig.~\ref{fig:model}a), his individual-level perception bias is $(1/6)/(1/3) = 0.5$, which means that he underestimates the size of the minority group by a factor of $0.5$. Consequently, he overestimates the size of his own, majority group in the overall network. In the heterophilic network (Fig.~\ref{fig:model}b), the perception bias of individual $i$ is $(4/6)/(1/3) = 2$, implying that he overestimates the size of the minority group by a  factor of $2$. At the group level, the majority group (blue) perceives the size of the minority group to be $4/18$ in the homophilic network and $10/18$ in the heterophilic network. Therefore, the majority group underestimates the size of the minority group by a factor of $(4/18)/(1/3) = 0.67$ in the homophilic network and overestimates it by a factor of $(10/18)/(1/3) = 1.67$ in the heterophilic network.

\subsection*{Survey of social perception biases}

To gain insight into the role of homophily and minority-group size in social perception biases, we conducted a survey with $N=101$ participants from the United States and $N=99$ participants from Germany, recruited from Amazon's Mechanical Turk, and $N=100$ South Korean participants recruited from the survey platform Tillion Panel. For American and German participants, we asked questions about 10 attributes, listed in Table~\ref{tb:survey_source}. The questions were taken from existing national surveys that provided objective frequencies of different attributes in the general population of each country. For Korean participants, we asked questions about seven of these attributes for which we had objective frequencies from national surveys (see Table~\ref{tb:survey_questions}). 

Participants answered three groups of questions. First, they answered questions about their own attributes (e.g., smoking). Second, they estimated the frequency of people with each attribute in their personal networks, defined as ``all adults you were in personal, face-to-face contact with at least twice this year.'' We used these answers to calculate the homophily in their personal networks (see Method for more details). For example, if a participant was a smoker and 70\% of her social contacts were smokers, the probability of a friendship link between this participant to a smoker is 70\%. We used this information to estimate the homophily parameter ($h$) for each individual's personal network using Eq.~\ref{eq:maaforh}. Homophily can vary from 0 (complete heterophily) to 1 (complete homophily). 

Third, participants estimated the frequency of people with a particular attribute in the general population of their country. We compared these estimates with the results from national surveys to calculate their social perception bias (see Method). The perception bias was calculated as a ratio of an individual's estimate and the objective frequency from national surveys. For example, if a participant reported that she believes that 60\% of the population smoke tobacco whereas the national survey suggested that 40\% do so, the perception bias was $60/40 = 1.5$. 

For each country, we analyzed perception biases separately for attributes that in that country are objectively found to be held by a small ($f_a < 0.2$), medium ($0.2 \leq f_a < 0.4$), or large ($0.4 \leq f_a < 0.5$) fraction of the minority group. For example, the attribute ``not having money for food'' is held by a small minority in all three countries (Table~\ref{tb:survey_questions}). In contrast, the attribute ``donating to charity'' is held by a large minority in Germany and a medium minority in South Korea. In the United States, this attribute is held by a majority, and its reverse, ``not donating to charity,'' is held by a large minority.


Fig.~\ref{fig:survey} shows the survey results for the United States and Germany (Fig.~\ref{fig:survey_korea} for the South Korea), with a different country in each row. Perception bias for the size of the minority group is shown separately for participants who belonged to the minority (left column) and majority (right column) groups. The value of the perception bias indicates how accurately each participant (each point in the plot) perceived the size of the minority group to be in the overall population. A perception bias of 1 means perfect accuracy, values above 1 indicate overestimation of the minority-group size, and values below 1 indicate underestimation of the minority-group size. We observed clear effects of the objective fraction of the minority group in the overall population ($f_a$) and of homophily of personal networks ($h$) on perception biases. As minority-group size in the overall population decreased, the size of the perception bias increased. Moreover, when homophily in personal networks was large ($h > 0.5$), minority participants overestimated and majority participants underestimated the size of the minority, resembling false consensus. In contrast, for low levels of homophily in personal networks ($h < 0.5$), we observed a much smaller false-consensus or even false-uniqueness tendencies for both minority and majority participants. Similar relationships between perception biases, homophily, and minority-group size were observed in all three countries.

The survey results provide evidence that homophily of personal networks and minority-group size partially explain perception biases regardless of cultural differences. They stress the necessity of considering the social structure in which an individual is embedded to understand how social perception biases are formed. In the next section we therefore explore whether these biases can be explained by a simple generative network model.

\subsection*{Generative network model with tunable homophily and minority-group size}
\label{sec:random}

To systematically study the relation between perception biases and network structure, we developed a network model that allows us to create scale-free networks with tunable homophily and minority-group sizes\cite{karimi2017visibility}. This network model is a variation of the Barab\'{a}si--Albert preferential attachment model with the addition of a homophily parameter $h$ (we call this model BA-homophily). The probability of an attachment of a newly arriving individual (node) $w$ to an existing node $v$ ($\phi_{wv}$) is proportional to the node's degree ($k_v$) and the homophily between the two nodes ($h_{wv}$); thus $\phi_{wv} \propto h_{wv}k_{v}$. 

Thus, the degree and the homophily parameter regulate the probability of connection between individuals who share the same attribute. The value of homophily ranges from 0 to 1. If homophily is low ($0 \leq h < 0.5$), nodes have a tendency to connect to nodes with opposite attribute values. As the homophily parameter increases above $0.5$, the probability of connection between nodes with the same attribute increases. In the case of extreme homophily ($ h = 1 $), nodes strictly connect to other nodes with the same attribute value and therefore two separate communities emerge. One advantage of using the BA-homophily model is that it generates networks with the scale-free degree distributions observed in many large-scale social networks. Another advantage is that it generates networks in which homophily can be symmetric or asymmetric. Given nodes with two attributes, $a$ and $b$, when homophily is symmetric the tendency to connect to nodes with the same attribute is the same for both groups, $h_{aa} = h_{bb} = h$. In this case one parameter is sufficient to generate the network. In the asymmetric homophily case, two homophily parameters are needed to regulate the connectivity probabilities for each group separately, thus  $h_{aa} \neq h_{bb}$.

\begin{figure*}[ht]
\centering
\begin{tabular}{l}\\
\includegraphics[width=0.8\linewidth]{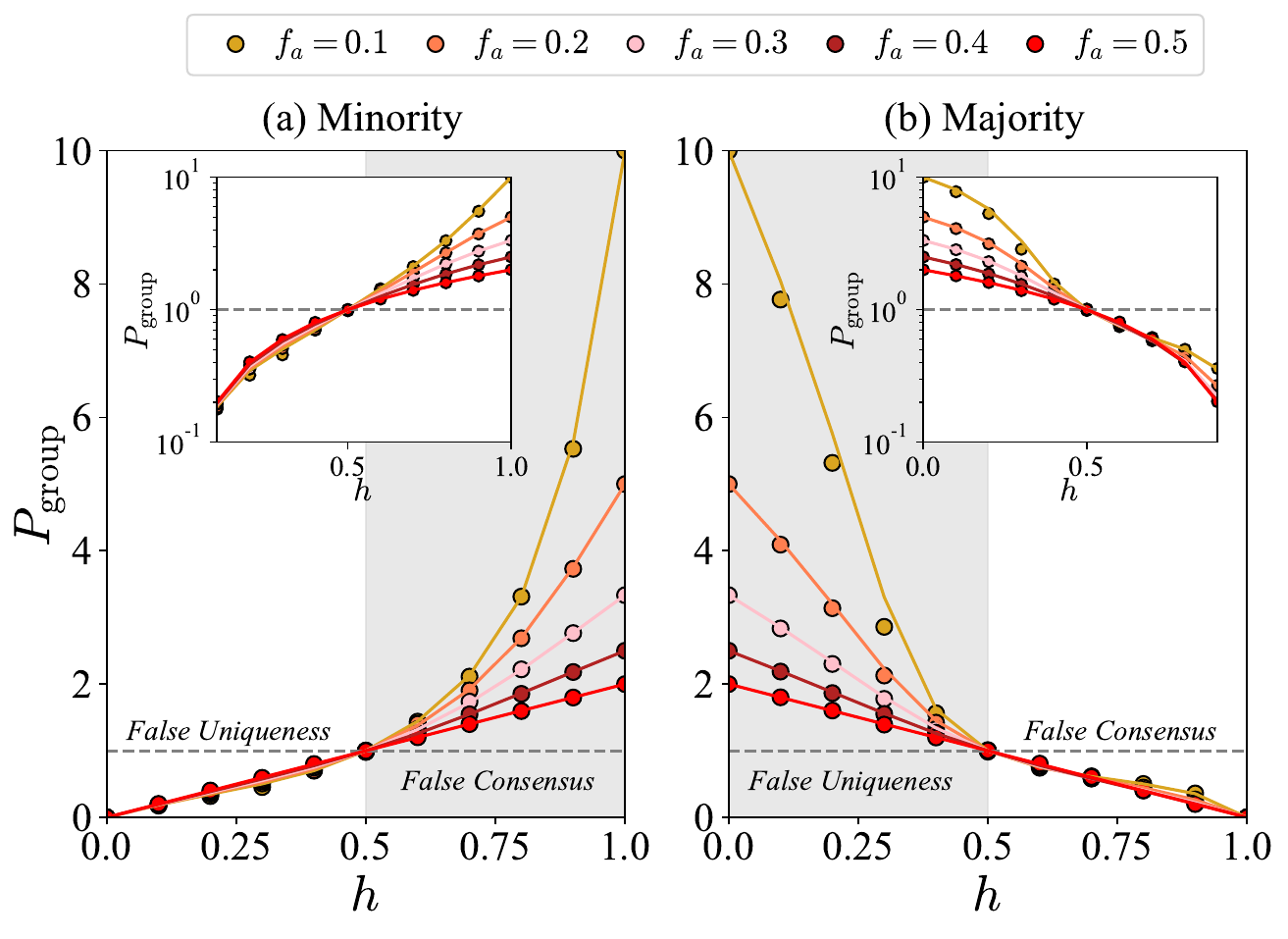}\\
\end{tabular}

\caption{
\textbf{Estimation of minority group size by (a) the minority group and (b) the majority group, as a function of homophily of personal networks ($\bm{h}$) and the minority-fraction ($\bm{f_a}$) in the overall network.}
Different colors refer to networks with different minority fractions ($f_a$). 
The analytic results are displayed as solid lines and numerical results as circles. In the heterophilic networks ($ 0 \leq h < 0.5$), the minority (a) underestimates its own size, and the majority (b) overestimates the size of the minority, resembling false uniqueness. In homophilic networks ($ 0.5 < h \leq 1$), the minority (a) overestimates its own size and the majority (b) underestimates the size of the minority, resembling false consensus. The insets show the same information on a log scale to make the amount of underestimation and overestimation comparable.
}
\label{fig:neterr}
\end{figure*}

Fig.~\ref{fig:neterr} depicts analytically derived biases in perceptions of minority-group size among individuals who themselves belong to a minority (Fig.~\ref{fig:neterr}a) or a majority (Fig.~\ref{fig:neterr}b) group, as a function of the true fraction of the minority in the overall network ($f_a$) and the homophily of individuals' personal networks ($h$). The solid lines show the analytic results (see Method) and the circles are numerical results obtained from the BA-homophily model. In heterophilic networks ($ 0 \leq h < 0.5$), perception biases resemble false uniqueness. The minority underestimates its own size, and the majority overestimates the size of the minority, the more so the smaller the minority group (smaller $f_a$). In homophilic networks ($ 0.5 <  h \leq 1$), perception biases resemble false consensus. The minority overestimates its own size (the more so the smaller the minority group), and the majority underestimates the size of the minority. Slight asymmetries between biases expected for minority and majority groups (see insets in Fig.~\ref{fig:neterr}) are due to the disproportionate number of links for two groups, affecting the results of Eq.~\ref{eq:pgroup_before}.

These analytic derivations can help us describe the functional form of the biases observed in the survey (Fig.~\ref{fig:survey}). As shown in Eq.~\ref{eq:pgroup_sort}, the minority's perception bias ($P^{a}_{\textrm{group}}$) is proportional to the density of links between minorities ($p_{aa}$), which increases with the homophily between minority nodes, that is, $h_{aa}$. Similarly, the majority's perception bias ($P^{b}_{\textrm{group}}$) is proportional to the density of intergroup links ($p_{ab}$), which decreases as the homophily ($h_{aa}$ and $h_{bb}$) increases. In addition, the relative sizes of minority and majority groups influence the growth rate of links for each group according to Eq.~\ref{eq:C} so that 
perception biases can increase (or decrease) nonlinearly with group size (see the Supplementary Materials). For instance, in the extreme homophily case with $h=1$, one gets $p_{aa}=p_{bb}=1$, while $p_{ab} = p_{ba} = 0$, leading to the minority's group-level perception bias of $1/f_a$. In sum, the proposed BA-homophily model and its analytic derivations facilitate systematic understanding of how network structure affects perception biases. 


While we find general agreement between the survey results and our BA-homophily model calculations, there are some differences that call for more detailed investigation in the future. One main difference is that in survey results (Fig.~\ref{fig:survey}) we observed perception bias $> 1$ in some cases when Fig.~\ref{fig:neterr} predicts it to be $< 1$. Specifically, this tends to happen for small minority sizes, when $h<0.5$ for minority and when $h>0.5$ for majority participants. A possible explanation that is in line with previous studies in social cognition is that people do not observe and report attribute frequencies in their samples (here, their personal networks) completely accurately, but with some random noise. This does not matter on average when minority size is relatively large as errors of over- and underestimation cancel out. But for small minority-group size the estimate cannot be lower than 0, meaning that the former errors (overestimated - true sample frequency) will be larger than the latter errors (true - underestimated sample frequency) and will not cancel out\cite{Juslin2007}. Hence, people's estimates of the frequency of attributes in their samples will be overestimated for small minority groups, which is what we observed in survey results. 

\subsection*{Social perception biases in real-world networks}
\label{sec:data}

The BA-homophily model offers a very simple representation of real-world networks. To examine possible social perception biases in the real world, we studied six empirical networks with various ranges of homophily and minority-group sizes. The network characteristics of the data sets are presented in Table~\ref{tab:stat}. The detailed descriptions of the data sets and references can be found in the Method section. These empirical networks have different structural characteristics and show different levels of homophily or heterophily with respect to one specific attribute (see Supplementary Materials). In five of the networks this attribute is gender (female or male), while in one - the American Physical Society (APS) network - the attribute is whether a paper belongs to the field of classical statistical mechanics or quantum statistical mechanics. 

To estimate homophily, we first assumed that homophily is symmetric in all networks. The symmetric homophily is equivalent to Newman's assortativity measure ($q$), which is the Pearson coefficient of correlation between pairs of linked nodes according to some attribute (e.g., race~\cite{newman2003mixing}). The Newman assortativity measure corresponds directly to the homophily parameter in our model when adjusted for the scale. In our model $h = 0$ means complete heterophily ($q = -1$), $h=0.5$ indicates no relationship between structure and attributes ($q = 0$), and $h = 1$ indicates complete homophily ($q=1$; see the Supplementary Materials). 

In reality, however, the tendency of groups to connect to other groups can be asymmetric. For example, it has been observed that in scientific collaborations, homophily among women is stronger than homophily among men \cite{jadidi2017gender}. Given the relationship between number of edges that run between nodes of the same group and homophily in Eqs.~\ref{eq:maaforh} and ~\ref{eq:asym_h}, we can estimate the asymmetric homophily, which differs for the minority ($h_{aa}$) and the majority ($ h_{bb}$) group (see Method). As we describe below, it turns out that asymmetric homophily has an important impact on the predictability of perception bias in empirical networks.

We used the measured homophily and minority-group size in the empirical networks (Table~\ref{tab:stat}) to generate synthetic networks with similar characteristics to those of the six empirical social networks. This enabled us to compare perception biases in empirical and synthetic networks and gain insights about the impact of homophily and minority-group size differences on possible individual- and group-level perception biases.

\begin{figure}[ht]
\includegraphics[width=0.98\linewidth]{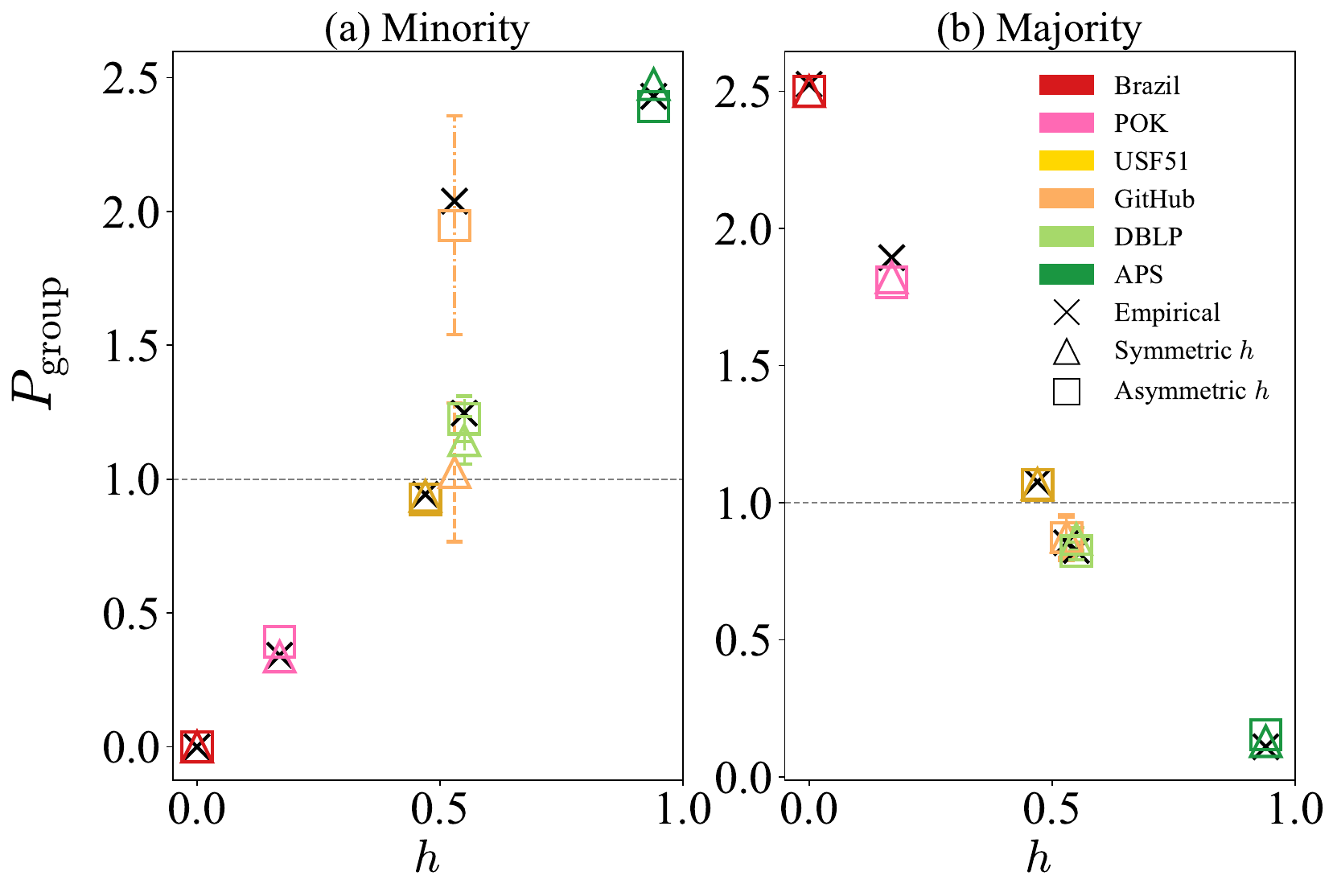} 
\caption{
\textbf{ Group-level social perception biases that could occur in six empirical social networks. The figure shows how accurately (a) the minority group and (b) the majority group might estimate the size of the minority group in real-world social networks with different levels of homophily.}
The symmetric homophily values of the empirical social networks are depicted on the \textit{x} axis. 
Homophily is measured between genders (female and male) except for the American Physical Society (APS) data where homophily is measured between different academic fields: classical statistical mechanics and quantum statistical mechanics. Empirical estimates of perception biases (crosses) are compared with estimates from the BA-homophily model assuming \textit{symmetric} (triangles) and \textit{asymmetric} (squares) homophily. For both types of homophily, the perception bias of the minority group increases as homophily increases in a network, and that of the majority group decreases as the homophily increases. The results of the BA-homophily model with asymmetric homophily are in excellent agreement with the empirical estimates, highlighting the importance of considering asymmetric homophily. The results of the BA-homophily model assuming symmetric homophily predict the trend well except for networks with high asymmetric homophily. The synthetic networks were generated with $N=2,000$ nodes and averaged over $20$ simulations. Standard deviations are shown if they are larger than a marker size.}
\label{fig:data_net}
\end{figure}

Fig. \ref{fig:data_net} shows group-level perception biases in the empirical networks that could occur if people's perceptions were based solely on the samples of information from their personal networks. Because further cognitive or motivational processes could affect the final perceptions, these estimates can be taken as a baseline level of biases that could occur without any additional psychological assumptions. The overall trends shown in Fig. \ref{fig:data_net} are in agreement with the results obtained from the survey and from the synthetic networks. In heterophilic networks, the minority group is likely to underestimate its own group size and the majority group is likely to overestimate the size of the minority. Conversely, in the homophilic networks, the minority group is likely to overestimate its own size and the majority group to underestimate the size of the minority. 

We can compare perception biases estimated directly from empirical networks (crosses in Fig. ~\ref{fig:data_net}) with those estimated from synthetic networks with similar homophily and minority-group size. In Fig.~\ref{fig:data_net}, triangles correspond to networks with symmetric homophily and squares to networks with asymmetric homophily. Although symmetric homophily traces empirically observed perception biases in most instances, it fails to capture the biases in the GitHub network, especially for the minority group. This network exhibits a higher level of asymmetric homophily compared to other networks (see Table~\ref{tab:stat}). When perception biases are estimated from a synthetic model that assumes asymmetric homophily, they approximate perception biases of both the minority and the majority groups very well in all networks. This suggests that asymmetric homophily plays an important role in shaping possible perception biases.

It is known that influential nodes in networks, usually identified by their high degree, can affect processes in networks such as opinion dynamics \cite{Aral2012}, social learning \cite{golub2010naive}, and wisdom of crowds~\cite{becker2017network}. To evaluate the impact of degree on shaping perception biases, we plotted individual perception biases, ${P}_{\textrm{indv.}}$, versus individual degree in Fig. S2 in supplementary. 
The distribution of individual perception biases estimated from the BA-homophily model mostly corresponds to the empirically estimated distribution. In addition, nodes with low degrees display a higher variation in perception biases compared to nodes with high degrees. The model does not explain all the variation observed in the empirical networks. This can be due to incomplete observations of all social contacts in real networks or to other processes that we did not consider in generating the model. However, the model can still predict the trend we observed in the empirical data, which would not be predicted assuming random connectivity among individuals. This analysis enables us to understand the impact of influential individuals on perception biases exhibited by other individuals in the network. 

\subsection*{Reducing social perception biases}


We investigated to what extent and under what structural conditions individuals can reduce their perception bias by considering the perceptions of their neighbors. We built on DeGroot's weighted belief formalization by aggregating an individual's own perception (\textit{ego}) with the averaged perceptions of the individual's direct neighbors (\textit{1-hop}) \cite{degroot1974reaching}. For simplicity, we assumed symmetric homophily in the BA-homophily model (the results for asymmetric homophily are in the Supplementary Materials). 

Fig. \ref{fig:hop} shows a comparison of the average perception bias ($\bar{P}_{\textrm{indv.}}$) for individuals who belong to the (a) minority and (b) majority and the weighted averages of their own perceptions and those of their 1-hop neighbors. The minority-group size is fixed to $0.2$. 
The results show that taking into account perceptions of 1-hop neighbors improves estimates of individuals in heterophilic networks (blue triangles are closer to the gray dashed line compared to orange circles in the log scale). The improvement in the perception is the result of nodes being more likely to be exposed to neighbors with opposing attributes. In homophilic networks, including neighbors' perceptions does not lead to a significant improvement because individuals are exposed to neighbors with similar attributes to their own.

This suggests that in homophilic networks, individuals cannot improve their perceptions by consulting their peers because they are similar to them and do not add new information that might not increase the accuracy of their estimates. However, in heterophilic networks, individuals benefit from considering their neighbors’ perceptions because they are not exposed only to other like-minded individuals but to more diverse perceptions. While the overall trend is not surprising, our results reveal how the accuracy of these estimates changes as a function of homophily.

\section*{Discussion}
The way people perceive their social networks influences their personal beliefs and behaviors and shapes their collective dynamics. However, many studies have documented biases in these social perceptions, including both false consensus and false uniqueness. Here we investigated to what extent both of these biases can be explained merely by the structure of the social networks in which individuals are embedded, without any assumptions about biased cognitive or motivational processes. 

Using empirical survey data, analytical investigation, and numerical simulations, we show that structural properties of personal networks strongly affect the samples people draw from the overall population. We find that biased samples alone can lead to apparently contradictory social perception biases such as false consensus and false uniqueness. While cognitive and motivational processes undoubtedly play an important role in the formation of social perceptions\cite{marks1987ten}, our analyses establish a baseline level of biases that can occur without assuming biased information processing\cite{fiedler2000beware,gigerenzer2012rethinking,Juslin2007,le2011rational,denrell2016information}.
 
Our results suggest that homophily impacts the accuracy of the estimates of individuals in both minority and majority groups. In homophilic networks, minority group tends to overestimate its own size, and majority group tends to underestimate the size of the minority.  In contrast, in heterophilic networks, minority group tends to underestimate its own size, and majority group overestimates the size of the minority. In other words, when homophily is high, both minority and majority groups tend to show social perception biases similar to false consensus, whereas when it is low, both group show biases similar to false uniqueness.

We further show that the relative sizes of the majority and minority groups influence perception biases. Specifically, the smaller the size of the minority, the higher the false consensus among members of the minority group and false uniqueness among members of the majority group. 

To explain the underlying structural mechanisms of perception biases in social networks, we developed a generative network model with tunable homophily and minority-group size. We find that predictions from this theoretical model correspond to empirical observations well, especially when we assume asymmetric rather than symmetric homophily. In addition, we show that perception biases can be reduced by aggregating individuals' perceptions with those of their direct neighbors, though only in heterophilic networks. In homophilic networks, these socially informed estimates do not lead to more accurate perceptions due to similarity of the nodes to their neighbors.

Most previous explanations of perception biases in social psychology did not include precise quantitative models of social environments. Many of them did not explicitly include the structure of social networks at all. Those that did, such as selective-exposure explanations, typically involved qualitative statements that people tend to socialize with similar others\cite{Ross1977,Sherman1983}. The network model described in the present paper enabled a thorough quantitative investigation of how different levels of homophily, its asymmetric nature, and different relative sizes of majority and minority groups, affect social perception biases. This investigation did not include a quantitative specification of the cognitive processes underlying people's sampling from their personal networks. Such specifications\cite{fiedler2000beware,pachur2013intuitive,le2011rational,galesic2018sampling} could be combined with the network model described here. 

Besides providing a theoretical account of perception biases, our results have practical implications for understanding real-world social phenomena. Given the importance of homophilic interactions in many aspects of social life ranging from health-related behavior~\cite{Centola2011} to group performance \cite{mollica2003racial} and social identity \cite{mehra1998margins}, it is crucial to consider obstacles to improving social perceptions. Perceptions of the frequency of different beliefs and behaviors in the overall population influence people's beliefs about what is normal and shape their own aspirations \cite{Festinger1954,suls2002social,Centola2011}. When people overestimate the prevalence of their own attributes in the overall population, they will be more likely to think that they are in line with social norms, and consequently, less likely to change. We found that small minorities with high homophily are especially likely to overestimate their actual frequency in the overall network. If such committed minorities become resistant to change, they can eventually influence the whole network\cite{mobilia2003does,mobilia2007role}, and when such minorities have erroneous views, the whole society could be worse off. Our results further suggest that a possible way to correct biases is to promote more communication with and reliance on neighbors' perceptions. However, this can be useful only in conjunction with promoting more diversity in people's personal networks. Promoting more communication in homophilic networks does not improve perception biases. 

This study is not without limitations. One strong assumption in our methodology is that one's perception is based solely on information sampled from one's direct neighborhood. In the real world, individuals can also rely on other sources such as news reports, polls, and general education. In addition, we observe differences between the results of our survey and numerical simulations that call for future investigation on the impact of minority-group size on perceptions. In particular, experimental group studies could lead to a better understanding of how different network properties affect perceptions.

In sum, this study shows that both over-and under-estimation of the frequency of own view can be explained by different levels of homophily, asymmetric nature of homophily, and size of the minority group. Integration and quantification of the biases provide a rather comprehensive picture of the baseline level of human perception biases. We hope that this paper offers insights into how to measure and reduce perception gaps between different groups and fuels more work on understanding the impact of network structure on individual and group perceptions of our social worlds.

\section*{Method}

\subsection*{BA-homophily model} 
\label{sec:model}

To gain insight into how network structure affects perception biases, we developed a network model that allows us to create scale-free networks with tunable homophily and minority-group size\cite{karimi2017visibility}. This network model is a variation of the Barab\'{a}si--Albert model with the addition of homophily parameter $h$. In this model, the probability that a newly introduced node $w$ connects to an existing node $v$ is denoted by $\phi_{wv}$ and it is proportional to the product of the degree of node $v$, $k_v$, and the homophily between $w$ and $v$ as follows:

\begin{equation}
\phi_{wv} = \frac{h_{wv}k_v}{\sum_{v\in\{G\},v \neq w}^{}h_{wv}k_v}.\\
\end{equation}
Here, $h_{wv}$ is the probability of connection between nodes $v$ and $w$. This is an intrinsic value that depends on the group membership of $v$ and $w$. $\{G\}$ is a set of nodes in a graph $G$. 

Before constructing the network, we specify two initial conditions: (i) the size of the minority group and (ii) the homophily parameter that regulates the probability of a connection between minority and minority individuals, majority and majority individuals, minority and majority individuals, and majority and minority individuals. Each arrival node continues the link formation process until it finds $m$ nodes to connect to. If it fails to do so, for example, in an extreme homophily condition, the node remains in the network as an isolated node. The parameter $m$ guarantees the lower bound of degree and in our model is set as 2.  Although this parameter is fixed for each node, the stochasticity of the model ensures the heterogeneity of the degree distribution.


\subsection*{Analytic derivation for group-level perception bias}
\label{sec:pgroup_analytics}

Let us refer to the minority as $a$ and the majority as $b$. $K_{a}(t)$ and $K_{b}(t)$ are the total number of degrees for each group of the minority and the majority at time $t$, respectively. 
At each time step, one node arrives and connects with $m$ existing nodes in the network. Therefore, the total degree of the growing network at time $t$ is $K(t) = K_{a}(t) + K_{b}(t) = 2mt$. In this model, the degree growth is linear for both groups. Denoting $C$ as the minority's degree growth factor, we have

\begin{equation}
    K_a(t) = Cmt, \quad
    K_b(t) = (2-C)mt.
\end{equation}
\label{eq:K}

The probability of a connection between two minority nodes is the product of their degree and homophily:

\begin{equation}
\label{eq:paa}
p_{aa} = \frac{h_{aa}K_a(t)}{h_{aa}K_a(t)+h_{ab} K_b(t)}=\frac{h_{aa}C}{h_{aa}C+h_{ab}(2-C)},
\end{equation}

where $h_{aa}$ is the homophily between minority nodes, and $h_{ab}$ is the homophily between a minority and a majority in a relation of $1-h_{aa}$. The connection probability from a minority to a majority is the complement of $p_{aa}$ as

\begin{equation}
\label{eq:pab}
p_{ab} = \frac{h_{ab}K_b(t)}{h_{aa}K_a(t) + h_{ab}K_b(t)}=\frac{h_{ab}(2-C)}{h_{aa}C+h_{ab}(2-C)}.
\end{equation}

Similar relationships can be found for the connection probability of majority to majority and majority to minority. 

To calculate the group perception bias ($P_{\textrm{group}}$), one needs to consider the number of inter- and intragroup edges as expressed in the definition of Eq.~\ref{eq:p_group}. Thus, the group perception bias for the minority and majority is as follows:

\begin{equation}\label{eq:pgroup_before}
\begin{split}
P^a_{\textrm{group}} = \frac{1}{f_a}\frac{2M_{aa}}{2M_{aa}+(M_{ab}+M_{ba})},\quad P^b_{\textrm{group}} = \frac{1}{f_a}\frac{M_{ab}+M_{ba}}{2M_{bb}+(M_{ab}+M_{ba})}.
\end{split}
\end{equation}

Here, $M$ represent the number of edges between different groups. For example, $M_{aa}$ is the number of edges between minority nodes. Note that we distinguish number of edges between minority and majority $M_{ab}$ and between majority and minority $M_{ba}$. These values are equivalent when homophily is symmetric but they are unequal when the homophily is asymmetric. Since each $M$ has a relation as a product of the total number of edges and the edge probability, such as $M_{aa} = mN_ap_{aa}$, we can reduce Eq.~\ref{eq:pgroup_before} to the following equations: 

\begin{equation}\label{eq:pgroup_sort}
\begin{split}
    P^a_{\textrm{group}} = \frac{1}{f_a}\frac{2p_{aa}}{2p_{aa} + p_{ab} + (N_b/N_a) p_{ba}},\quad P^b_{\textrm{group}} = \frac{1}{f_a}\frac{(N_a/N_b) p_{ab}+p_{ba}}{2p_{bb}+(N_a/N_b) p_{ab} + p_{ba}},
\end{split}
\end{equation}

where $N_a$ and $N_b$ represent the number of nodes in each group. The analytic derivations are intuitive and well explained by the numerical results (solid lines in Fig.~\ref{fig:neterr}). For example, when $f_a = 0.5$ in extreme homophily ($h=1.0$) with the degree growth $C = 1$ (a symmetric homophily condition), $P^a_{\textrm{group}}  = 2$ from Eq.~\ref{eq:pgroup_sort}, and it matches well with the numerical result in Fig.~\ref{fig:neterr}a. Note that the growth parameter $C$ is a polynomial function and its relation to homophily is shown in the Supplementary Materials.

\subsection*{Measuring homophily in empirical networks}

From the linear degree growth shown in Eq.~\ref{eq_sp:dka}, we can derive the relation between the degree growth $C$ and the inter- and intralink probabilities $p_{aa}, p_{ab}$, and $p_{bb}$ (see the Supplementary Materials). Thus, 

\begin{equation}
    C = f_a(1 + p_{aa}) + f_b p_{ba}.
\label{eq:C}
\end{equation}

In empirical networks we know the edge density for the minority ($r_{aa} = M_{aa}/M$), between groups ($r_{ab}= M_{ab}/M$), and for the majority ($r_{bb}= M_{bb}/M$). Thus, the probability of connection within a group can be written as $r_{aa} = f_a p_{aa}$ and $r_{bb} = f_b p_{bb}$. 

Connection probabilities are defined as follows:

\begin{equation}
\label{eq:maaforh}
\begin{split}
p_{aa} &= \frac{h_{aa}C}{h_{aa}C+h_{ab}(2-C)},\quad p_{bb} &= \frac{h_{bb}(2-C)}{h_{bb}(2-C)+h_{ba}C}.
\end{split}
\end{equation}

From Eq.~\ref{eq:maaforh} and the relation between $r_{aa}$ and $p_{aa}$ (or $r_{bb}$ and $p_{bb}$), we can derive the  empirical homophily by using edge density $r_{aa},r_{bb}$ as follows:

\begin{equation}
\label{eq:asym_h}
\begin{split}
h_{aa} &= \frac{r_{aa}(2-C)}{f_aC+2r_{aa}(1-C)},\quad h_{bb} &= \frac{r_{bb}C}{f_b(2-C)-2r_{bb}(1-C)}.
\end{split}
\end{equation}

This calculations allow to estimate the homophily from the empirical networks assuming that the BA-homophily model is a valid model of a social network. The homophily can be symmetric or asymmetric in its nature. In order to understand the role of asymmetricit homophily in perception biases, we first assume all empirical networks have symmetric homophily. We approximate symmetric homophily by projecting the homophily to Newman's assortativity measure ($q$) (see Supplementary Materials and Fig.~\ref{fig:newman_assort}).

\subsection*{Empirical networks}
\label{sec:empi_networks}
The first network, Brazilian network, captures sexual contact between sex workers and sex buyers  \cite{rocha2011simulated}. The network consists of $16,730$ nodes and $39,044$ edges. There are $10,106$ sex workers and  $6,624$ sex buyers (minority-group size $f_a = 0.4$). In this network, no edges among members of the same group exist resulting in the Newman's assortativity ($q=-1$), and consequently, the network is purely heterophilic ($h=0$). 

The second network is an online Swedish dating network from PussOKram.com (POK)~\cite{holmepok}. This network contains $29,341$ nodes with strong heterophily ($h=0.17, q=-0.65$). Given the high bipartivity of the network, we are able to infer the group of nodes using the max-cut greedy algorithm. The results are in good agreement with the bipartivity reported ~\cite{holmebipart}. Since the group definition is arbitrary, we label the nodes based on their relative group size as minority gender and majority gender. Here, the fraction of the minority in the network is $0.44$. 

The third network is a Facebook network of a university in the United States (USF51)~\cite{facebook}. The network is composed of 6,253 nodes and includes information about individuals' gender. In this network male students are in the minority, occupying $42\%$ of the network, and the network exhibits a small heterophily~\cite{facebook} ($q=-0.06, h=0.48$).

The fourth network is extracted from the collaborative programming environment GitHub. The network is a snapshot of the community (extracted August 4, 2015) that includes information about the first name and family name of the programmers. We used the first name and family name to infer the gender of the programmers \cite{karimi2016inferring}. After we removed ambiguous names, the network consisted of $120,338$ men and $7,330$ women. Here, women belong to the minority group and represent only about $6\%$ of the population. The network displays a moderately symmetric gender homophily of $0.53$ ($q=0.07$).

The fifth network depicts scientific collaborations in computer science and is extracted from  Digital Bibliography \& Library Project's website (DBLP) \cite{DBLP}. We used a new method that combines names and images to infer the gender of the scientists with high accuracy \cite{karimi2016inferring}. We used a 4-year snapshot for the network. After we filtered out ambiguous names, the resulting network included $280,200$ scientists and $750,601$ edges (paper coauthorships) with $63,356$ female scientists and $216,844$ male scientists. This network shows a moderate level of symmetric homophily ($h=0.55$ and $q=0.1$). 

The last network is a scientific citation network of the American Physical Society (APS). Citation networks depict the extent of attention to communities in different scientific fields. We used the Physics and Astronomy Classification Scheme (PACS) identifier to select papers on the same topics. Here, we chose statistical physics, thermodynamics, and nonlinear dynamical systems subfields (PACS = 05). Within a specific subfield, there are many subtopics that form communities of various sizes. To make the data comparable with our model, we chose two relevant subtopics, namely, Classical Statistical Mechanics (CSM) and Quantum Statistical Mechanics (QSM). 
The resulting network consists of $1,853$ scientific papers and $3,627$ citation links. Among nodes, $696$ are in the minority and $1,157$ in the majority. Here, the minority group in these two subtopics is CSM ($f_a = 0.37$). This network shows the highest homophily compared to the other empirical data sets ($h=0.94$ and $q=0.87$). 

\subsection*{Survey study}

For all countries (the Unite State, Germany, and South Korea), we restricted participant age to at least 18 years. Regarding gender, $61.4\%$ of the U.S. participants, $85.9\%$ of the German participants, and $50.0\%$ of the Korean participants were male. The age distribution in the United States was 18--30 years: $34.6\%$, 31--40 years: $36.6\%$, 41--50 years: $9.8\%$, 50+ years: $19.0\%$; in Germany it was 18--30 years: $61.4\%$, 31--40 years: $26.3\%$, 41--50 years: $6.0\%$, 50+ years: $6.3\%$; and in South Korea it was 18--30 years: $26.0\%$, 31--40 years: $26.0\%$, 41--50 years: $24.0\%$, 50+ years: $24.0\%$. 

Participants were asked questions about their own attributes, the frequency of these attributes in their personal networks, and their frequency in the overall population of their country. Question texts and objective sizes of minority and majority groups in the overall populations were taken from publicly available results of large national surveys conducted in each country.
Details are provided in Table~\ref{tb:survey_source}. U.S. and German participants were asked about 10 attributes and Korean participants about seven of those attributes for which we could find objective population data.

We estimated the homophily of participants' personal networks on the basis of their reports of the size of minority and majority groups in their social circles and following an approach similar to that of Coleman~\cite{coleman1958}. Each survey participant reported the fraction of his or her personal network (or social circle) who have a specific attribute. For example, a participant who does not smoke might have estimated that $80\%$ of her social circle are nonsmokers. We used this fraction to calculate the probability that any two nonsmokers in her social circle are connected. As a complementary relation of connection between attributes, we furthermore used the fraction of smokers in her social circle---$20\%$---to calculate the probability that any nonsmoker and smoker are connected.
These probabilities are equivalent to $p_{aa}$ or $p_{bb}$, and their complementaries correspond to $p_{ab}$ or $p_{ba}$ in the BA-homophily model. Using Eqs.~\ref{eq:C} and ~\ref{eq:maaforh} we can calculate the homophily $h_{aa}$ (or $h_{bb}$) of each participant's personal network. In addition, we can evaluate $h_{ab}$ and $h_{ba}$ using the relations $h_{ab}=1-h_{aa}$ and $h_{ba}=1-h_{bb}$. 

To study the effect of minority-group size, we analyzed results separately for attributes for which minority group size in a particular country was small ($f_a < 0.2$), medium ($ 0.2 \leq f_a < 0.4$), and large ($0.4 \leq f_a < 0.5$). For example, small minority attributes in the U.S. are no money for food, experienced theft, and smoking, because the objective frequency of these attributes in the general U.S. population is smaller than $0.2$ (see Table~\ref{tb:survey_questions}). We measured participants' individual perception bias by dividing their estimate of minority-group size in the general population by the objective minority-group size obtained from national surveys, according to Eq.~\ref{eq:p_indiv}.



\clearpage
\textbf{Acknowledgments:} We thank Kristi Winters, Julian Kohne, Petter Holme, and Henrik Olsson for insightful conversations. \textbf{Funding:} 
We thank the Complex Systems Society's Bridge Fund and GESIS for funding E. Lee's research visit. E.L. was supported by the Basic Science Research Program through the National Research Foundation of Korea (NRF) funded by the Ministry of Science and ICT with grant no. NRF-2017R1A2B2005957. H.-H.J. acknowledges financial support from the Basic Science Research Program through an NRF grant funded by the Ministry of Education (2015R1D1A1A01058958). M.G. acknowledges financial support from National Science Foundation grants no. 1745154 and 1757211, and United States Department of Agriculture,
National Institute of Food and Agriculture grant no. 2018-67023-27677. \textbf{Author contributions:} All authors conceived the project, developed the argument, and wrote the paper. E.L. and F.K. conducted the experiments and wrote the code. E.L. and M.G. conducted the surveys and its analysis. \textbf{Competing interests:} The authors declare that they have no competing interests. \textbf{Data and materials availability:} The six empirical data and the codes for implementing the analysis can be found online at https://github.com/frbkrm/NtwPerceptionBias. The survey data can be obtained from the authors upon request.

\clearpage
\section*{Figures and Tables}

\smallskip

\begin{table}[ht]
\centering

\begin{tabular}{|l|l|l|l|l|l|}
\cline{1-6}

\begin{tabular}[c]{@{}l@{}} \textbf{Data} \end{tabular} & 

\begin{tabular}[c]{@{}l@{}} \textbf{Number of nodes} \end{tabular}&

\begin{tabular}[c]{@{}l@{}} \textbf{Minority} \end{tabular}&

\begin{tabular}[c]{@{}l@{}} \textbf{Majority} \end{tabular} & 
 
\begin{tabular}[c]{@{}l@{}} \textbf{Symmetric \textit{h}} 
\end{tabular} &  

\begin{tabular}[c]{@{}l@{}} \textbf{Asymmetric \textit{h}} 
\\(minority, majority)\end{tabular}  \\

\cline{1-6}
\textbf{Brazil}&  
\begin{tabular}[c]{@{}l@{}}
16,730\end{tabular} &
\begin{tabular}[c]{@{}l@{}}Sex workers\\ 6,624 (40\%)\end{tabular} & \begin{tabular}[c]{@{}l@{}}Sex buyers \\ 10,106\end{tabular} &  
\begin{tabular}[c]{@{}l@{}}0.0 \end{tabular} & \begin{tabular}[c]{@{}l@{}}0, 0 \end{tabular}\\ 

\cline{1-6}
\textbf{POK}   & \begin{tabular}[c]{@{}l@{}}
29,341\end{tabular} &\begin{tabular}[c]{@{}l@{}}Minority gender\\ 12,868  (44\%)\end{tabular}        & 
\begin{tabular}[c]{@{}l@{}}Majority gender\\ 16,473\end{tabular} & 
\begin{tabular}[c]{@{}l@{}}0.17 \end{tabular} & \begin{tabular}[c]{@{}l@{}}0.2, 0.17 \end{tabular}\\  

\cline{1-6}
\textbf{USF51} & \begin{tabular}[c]{@{}l@{}}
6,253\end{tabular} & 
\begin{tabular}[c]{@{}l@{}}Male\\ 2,626 (42\%)\end{tabular}          & \begin{tabular}[c]{@{}l@{}}Female\\ 3,627\end{tabular} &
\begin{tabular}[c]{@{}l@{}}0.47 \end{tabular} & \begin{tabular}[c]{@{}l@{}}0.48, 0.47 \end{tabular}\\

\cline{1-6}
\textbf{GitHub}   & \begin{tabular}[c]{@{}l@{}}
127,668\end{tabular} &\begin{tabular}[c]{@{}l@{}}Female\\ 7,330 (5.7\%)\end{tabular}      & 
\begin{tabular}[c]{@{}l@{}}Male\\ 120,338\end{tabular} & \begin{tabular}[c]{@{}l@{}}0.53 \end{tabular} & \begin{tabular}[c]{@{}l@{}}0.69, 0.54 \end{tabular}\\ 

\cline{1-6}
\textbf{DBLP}     &\begin{tabular}[c]{@{}l@{}}
280,200\end{tabular} & \begin{tabular}[c]{@{}l@{}}Female\\ 63,356 (22\%)\end{tabular}     & 
\begin{tabular}[c]{@{}l@{}}Male\\ 216,844\end{tabular}  & \begin{tabular}[c]{@{}l@{}}0.55 \end{tabular} & \begin{tabular}[c]{@{}l@{}}0.57, 0.56 \end{tabular}\\ 

\cline{1-6}
\textbf{APS}      &\begin{tabular}[c]{@{}l@{}}
1,853\end{tabular} & \begin{tabular}[c]{@{}l@{}}CSM\\ 695 (37\%)\end{tabular}          & 
\begin{tabular}[c]{@{}l@{}}QSM\\ 1,158\end{tabular}    &
\begin{tabular}[c]{@{}l@{}}0.94 \end{tabular} & \begin{tabular}[c]{@{}l@{}}0.88, 1.0 \end{tabular}\\ 

\cline{1-6}
\end{tabular}

\caption{\textbf{Characteristics of the empirical networks.} Shown are sizes (number of nodes) of the minority and majority groups regarding a gender attribute (for Brazil, POK, USF51, GitHub, DBLP) and an academic field (for APS), as well as the values of symmetric and asymmetric homophily. Values are rounded to the second decimal point. \label{tab:stat}}

\end{table}

\begin{figure*}[ht]
\centering
\includegraphics[width=0.95\linewidth]{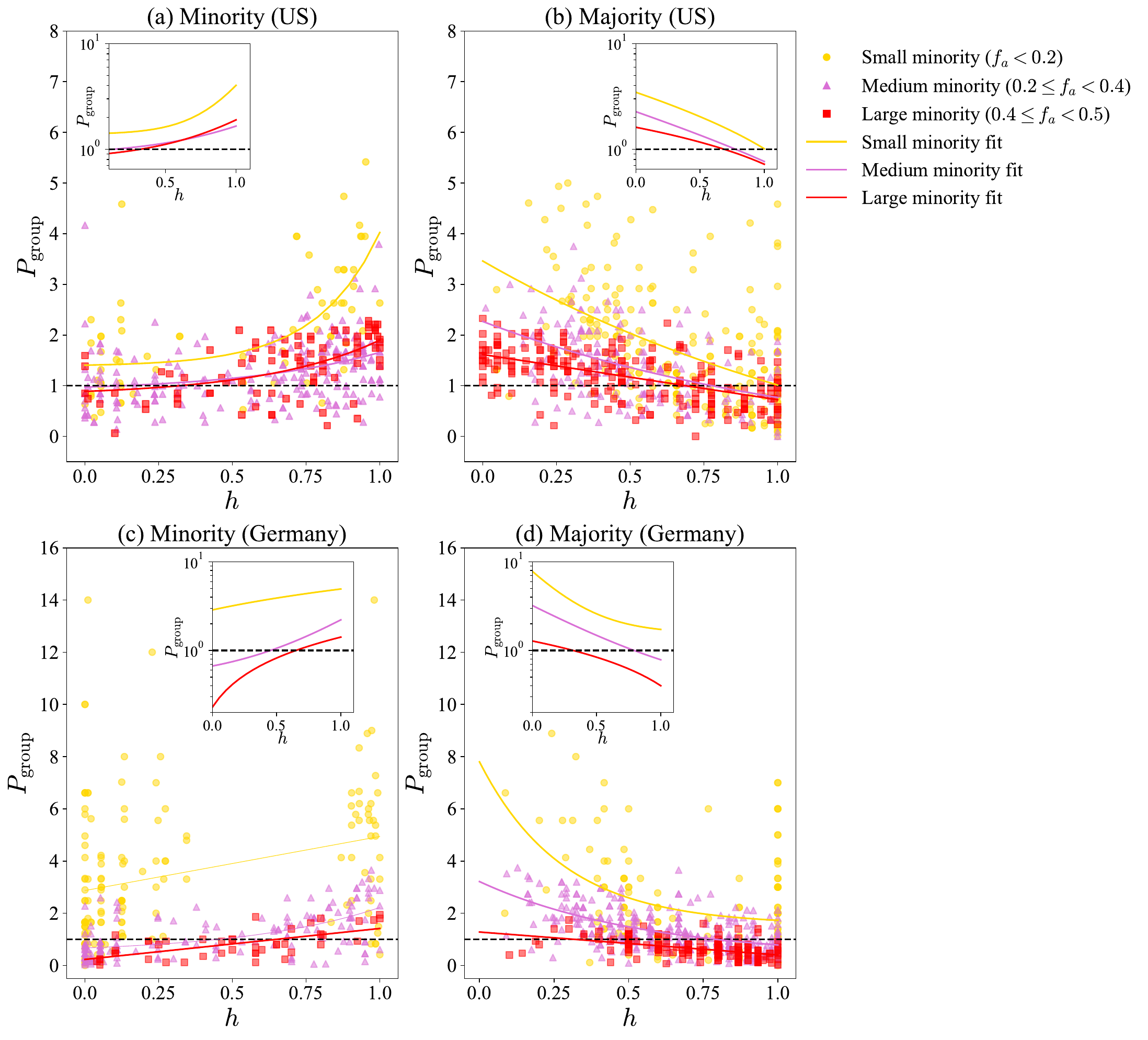}\\
\caption{\textbf{Bias in perception of minority-group size, for participants whose personal networks exhibit different levels of homophily ($\bm{h}$), and for attributes held by a  small, medium, or large minority group in a given country.} Rows shows results from different countries: United States (top) and Germany (bottom). Different colors distinguish perception biases for attributes that in a given country are held by a small ($f_a < 0.2$), medium ($0.2 \leq f_a < 0.4$), or large ($0.4 \leq f_a < 0.5$) minority group. Each data point represents the perception bias of a group where one individual involved for an attribute. Group level perception bias is calculated as a ratio of the perceived size of the minority group (obtained in our surveys) and the objective minority size (obtained from national surveys). Panels in the left column show perception biases of the minority, and panels in the right column show perception biases of the majority. The solid line is the curve fit of the data. The insets show fitted trends on a log scale to make the amount of underestimation and overestimation comparable. Homophily ($h$) is estimated from participants' reports about the minority-group fraction in their personal networks by mean-field assumption (see Supplementary Materials).}\label{fig:survey}
\end{figure*}

\clearpage

\clearpage

\clearpage

\begin{figure*}[ht]
\begin{center}
\includegraphics[width=0.85\linewidth]{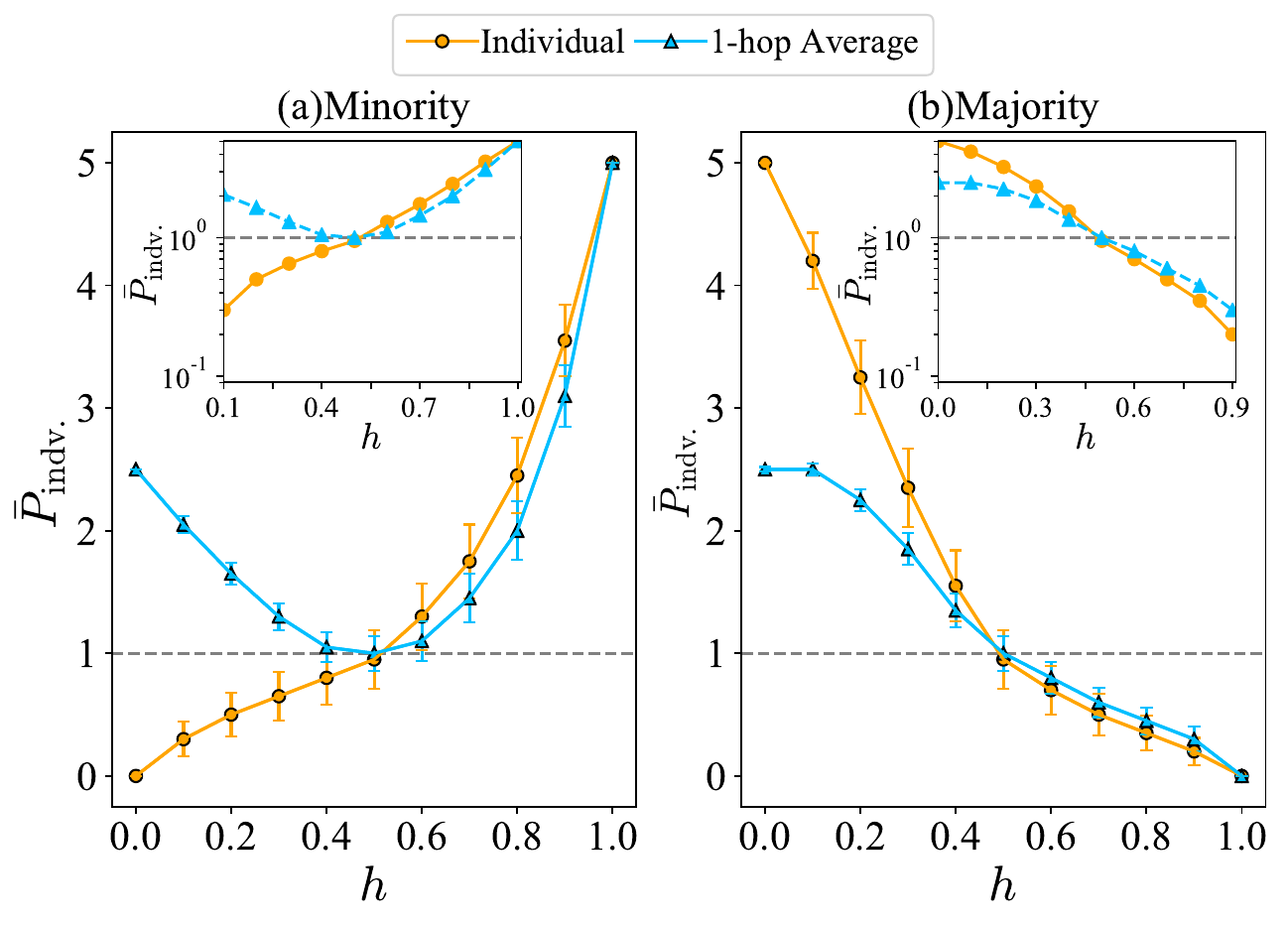}
\caption{
\textbf{Social perception biases for individual nodes and for the weighted average of perceptions of individual nodes and their 1-hop neighbors}. 
Insets show the same results in log scale. Orange lines are calculated from Eq.~\ref{eq:p_indiv} and averaged over all nodes in the group. Blue lines show the perception bias of the weighted average of perceptions of individual nodes and their direct neighbors (1-hop). We assume symmetric homophily, minority fraction of $0.2$, and networks with $2,000$ nodes. Results, averaged over 50 runs, show that perceptions of both minority and majority groups become slightly more accurate when taking into account their neighbors, but only in the heterophilic networks (in insets, blue triangles are closer than orange dots to the gray dashed line denoting less bias).} 
\label{fig:hop}
\end{center}
\end{figure*}


\clearpage
\section*{Supplementary Materials}

\renewcommand{\thefigure}{S\arabic{figure}}
\setcounter{figure}{0}
\renewcommand{\theequation}{S\arabic{equation}}
\setcounter{equation}{0}
\renewcommand{\thetable}{S\arabic{table}}
\setcounter{table}{0}

\subsubsectionfont{\normalfont\normalsize}

\section*{1. Survey questions and general population data}
\captionsetup{skip=0pt}
\begin{table}[!htp]
\small
\begin{center}
\begin{tabular}{ |p{3.0cm}|p{6.3cm}|p{1.7cm}|p{1.7cm}|p{1.8cm}| }
\hline
 \multicolumn{1}{|c|}{\multirow{2}{*}{\textbf{Characteristic}}}
 & \multicolumn{1}{c|}{\multirow{2}{*}{\textbf{Question text}}}
 & \multicolumn{3}{c|}{\textbf{\makecell{Original source of the question text \\ and general population data}}}\\
  \cline{3-5}
 &
 & \makecell{U.S.} 
 & \makecell{Germany} 
 & \makecell{S. Korea}\\  
 \hline
 \makecell[l]{1. Not having money\\ for food} & \makecell[l]{Have there been times in the past 12 months \\when you did not have enough money\\ to buy food you or your family needed? \\(a)Yes–(b)No}  & \makecell{Gallup\\ World Poll\\(2010)} & \makecell{Gallup\\ World Poll\\(2010)} &  \makecell{\small{Social}\\\small{Integration} \\\small{Status} \\\small{Survey}\\\small{(2011)}} \\\hline
 
 \makecell[l]{2. Donating to charity} & \makecell[l]{In the past month, have you donated \\money to a charity? (a)Yes–(b)No} & \makecell{Gallup\\World Poll\\(2010)} & \makecell{Gallup\\World Poll\\(2010)} & \makecell{\small{Social}\\\small{Survey of} \\\small{Welfare}\\\small{(2017)}}\\\hline
 
 \makecell[l]{3. Experiencing theft} & \makecell[l]{Within the past 12 months,\\ have you had money or property stolen\\ from you or another household member?\\ (a)Yes–(b)No} & \makecell{Gallup\\ World Poll\\(2010)} & \makecell{Gallup\\ World Poll\\(2010)} & \makecell{--} \\\hline 
 
 \makecell[l]{4. Religion importance} & \makecell[l]{Is religion an important part of your\\ daily life? (a)Yes–(b)No} & \makecell{Gallup\\ World Poll\\(2010)} & \makecell{Gallup\\ World Poll\\(2010)} & \makecell{\small{Religion}\\\small{of Koreans}\\\small{by Gallup}\\\small{(1984--2014)}}\\\hline 
 
 \makecell[l]{5. Worship attendance} & \makecell[l]{Have you attended a place of worship or \\ a religious service within\\ the past 7 days? (a)Yes–(b)No\\} & \makecell{Gallup\\ World Poll\\(2010)} & \makecell{Gallup\\ World Poll\\(2010)} & \makecell{\small{Religion}\\\small{of Koreans}\\\small{by Gallup}\\\small{(1984--2014)}}\\\hline 
 
 \makecell[l]{6. God and morality} & \makecell[l]{Which one of these comes closer to your \\opinion?\\
(a) It is not necessary to believe in God\\ in order to be moral and have good values.\\(b) It is necessary to believe in God\\ in order to be moral and have good values.
} & \makecell{Pew (2011)} &\makecell{Pew (2011)} & \makecell{--}\\\hline

 \makecell[l]{7. Belief in a god} & \makecell[l]{Do you believe in god or a supreme being?\\(a)Yes–(b)No} & \makecell{Pew (2011)} & \makecell{Pew (2011)} &\makecell{\small{Religion}\\\small{of Koreans}\\\small{by Gallup}\\\small{(1984--2014)}}\\\hline 
 
 \makecell[l]{8. Smoking} & \makecell[l]{These days, are you smoking any tobacco \\product at least once a day? \\(Tobacco smoking includes cigarettes,\\ cigars, pipes, and any other form of smoked \\tobacco).(a)Yes–(b)No} & \makecell{\small{World}\\\small{Health}\\\small{Organization}\\(2010)} & \makecell{\small{World}\\\small{Health}\\\small{Organization}\\(2010)} &\makecell{National\\ Nutrition\\Survey\\(2016)} \\\hline 
 
 \makecell[l]{9. Military force} & \makecell[l]{Do you agree that it is sometimes necessary \\to use military force to maintain order\\ in the world? (a)Yes–(b)No} & \makecell{Pew (2011)} & \makecell{Pew (2011)} &  \makecell{--} \\\hline
 
 \makecell[l]{10. Homosexuality\\acceptance} & \makecell[l]{Which one of these comes closer to your\\ opinion?\\
(a) Homosexuality is a way of life that should\\ be accepted by society.\\(b) Homosexuality is a way of life that should \\not be accepted by society.
} & \makecell{Pew (2011)} & \makecell{Pew (2011)} &\makecell{Gallup\\Daily\\ Opinion\\(2017)}\\
 \hline
\end{tabular}
\end{center}
\caption{\textbf{Survey questions about different attributes.} Survey questions asking about different attributes in the United States, Germany, and South Korea, including the source of question texts and the objective data for minority and majority group sizes in the general population.}\label{tb:survey_source}
\end{table}
\clearpage

\begin{table}[ht]
\begin{center}
\begin{tabular}{ |c|c|c|c|c|c|c| }
 \hline

 \hline
 \multirow{2}{*}{\textbf{Characteristic}}
 & \multicolumn{2}{c|}{U.S. (\%)} & \multicolumn{2}{c|}{Germany (\%)}
 & \multicolumn{2}{c|}{S. Korea (\%)}\\  
 \cline{2-7}
 & (a)
 & (b)
 & (a)
 & (b)
 & (a)
 & (b)\\
 \hline
 \makecell{1. Not having money for food}  & 19 & 81 & 5 & 95 & 3 & 97 \\ 
 \makecell{2. Donating to charity} & 57 & 43 & 43 & 57 & 26.7 & 73.7\\ 
 3. Experiencing theft & 12 & 88 & 9 & 91 & -- & -- \\ 
 4. Religion importance & 70 & 30 & 27 & 73 & 52 & 48\\ 
 5. Worship attendance & 53 & 47 & 33 & 67 & 44 & 56\\ 
 6. God and morality & 47 & 53 & 33 & 67 & -- & --\\ 
 7. Belief in a god & 64 & 36  & 38 & 62 & 39 & 61\\ 
 8. Smoking & 15.2 & 85 & 21.9 & 78 & 23.9 & 76.1\\ 
 9. Military force & 76.5 & 24 & 50 & 50 & -- & -- \\ 
 10. Homosexuality & 35.5 & 64 & 12.1 & 87.9 & 34 & 58\\ 
 \hline

\end{tabular}
\end{center}
\caption{\textbf{Objective size of minority and majority group for each attribute.} Objective data for minority and majority group sizes for each attribute in the general populations of the United States, Germany, and South Korea. Texts of answers (a) and (b) correspond to those listed in the Table~\ref{tb:survey_source} for each answer.}\label{tb:survey_questions}
\end{table}
\clearpage
\section*{2. Survey results for South Korea}

Survey results for South Korea show similar pattern as those for the United States and Germany, shown in Fig.~\ref{fig:survey}. In the heterophilic networks $(0 \leq h < 0.5)$, the minority (a) underestimates its own size, and the majority (b) overestimates the size of the minority, resembling false uniqueness. In homophilic networks $(0.5 < h \leq 1)$, the minority (a) overestimates its own size and the majority (b) underestimates the size of the minority, resembling false consensus.
We calculated the group level perception bias as a relative measure: the perceived size of the minority group divided by the objective minority size obtained in national surveys.
The right inset displays the same information on a log scale to make the amount of underestimation and overestimation comparable. The left inset shows the perception bias of small minority, whose bias ranges wider than for other groups. In general, as minority-group size becomes smaller, social perception biases increase. To fit the survey results, we used the curve fit.

\begin{figure*}[ht]
\centering

\includegraphics[width=0.96\linewidth]{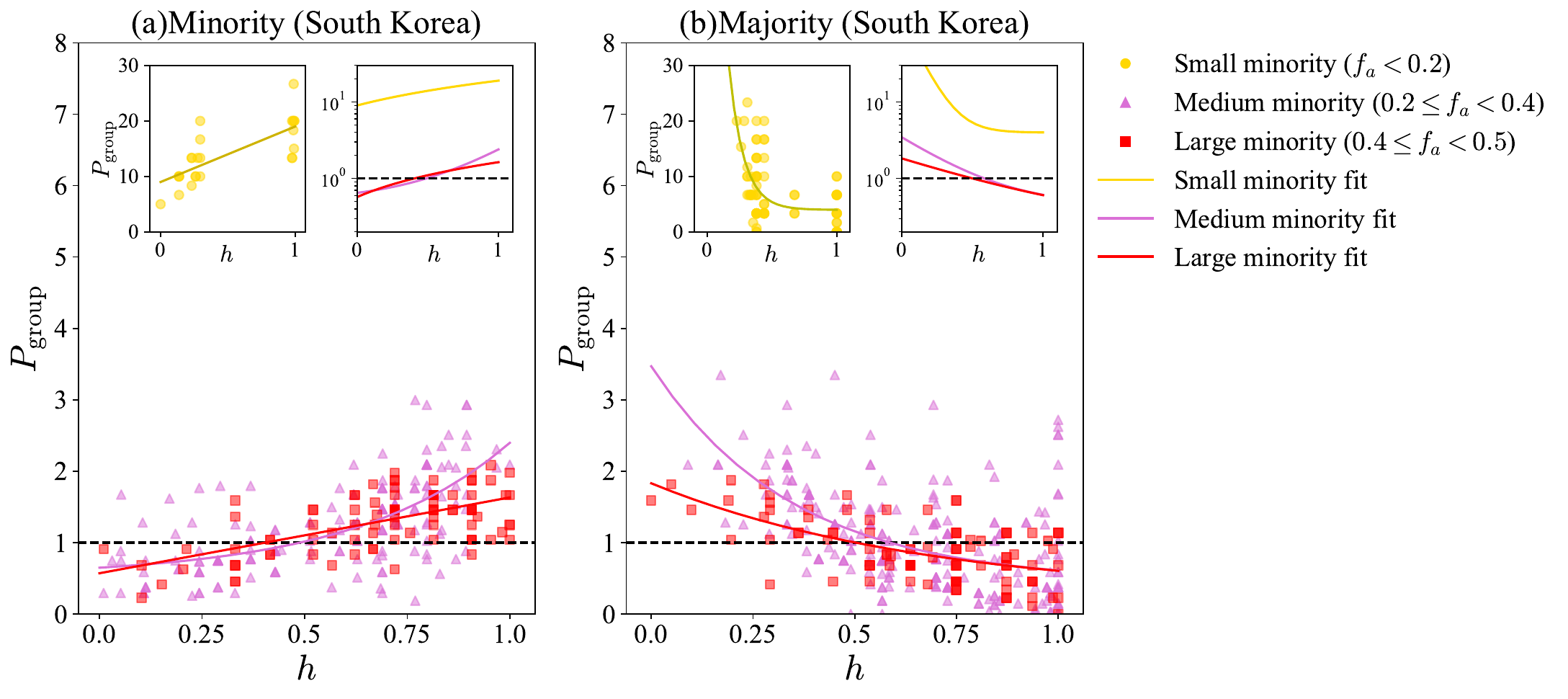}

\caption{
\textbf{Bias in perception of minority-group size, for participants whose personal networks exhibit different levels of homophily ($\bm{h}$), and for attributes held by a small, medium, or large minority group in South Korea.} 
Different colors distinguish perception biases for attributes that in South Korea are held by a small($f_a<0.2$), medium ($0.2\leq f_a < 0.4$), or large ($0.4\leq f_a < 0.5$) minority group. Each data point represents the perception bias of a group where one individual involved for an attribute. The left inset shows the bias of small minority group, and the right inset displays the same bias on a log scale to make the amount of underestimation and overestimation comparable.
The solid lines with colors for the group sizes are drawn by the curve fit.}
\label{fig:survey_korea}
\end{figure*}

\clearpage


\section*{\textbf{2. Distribution of individual-level perception bias as a function of degree}}

\begin{figure*}[ht]
\centering
\includegraphics[width=0.95\linewidth]{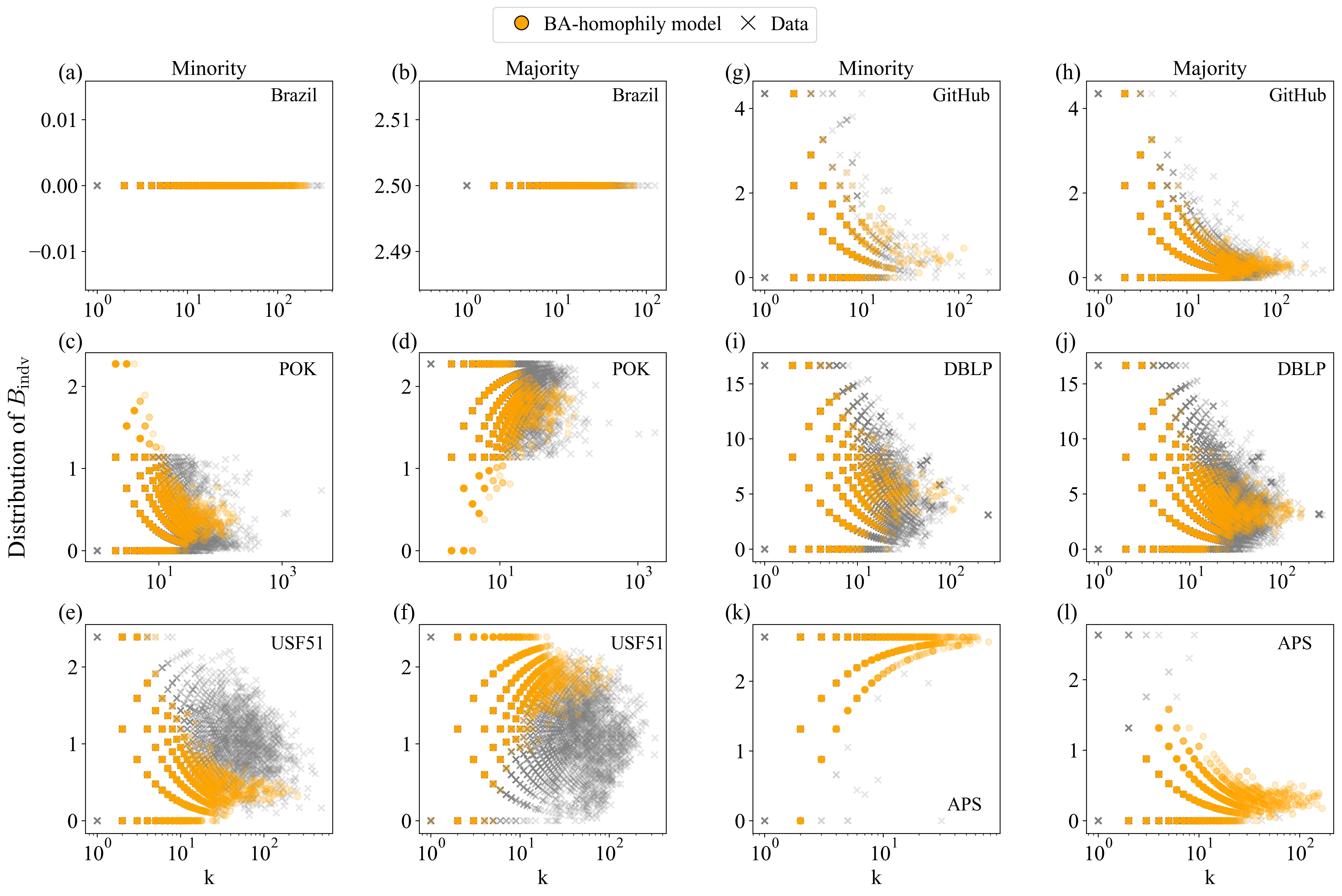}
\caption{
\textbf{Distribution of individual-level perception bias ($\bm{P}_{\textrm{indv.}}$) as a function of degree ($\bm{k}$).} Each row represents one empirical network. The left two columns show the distribution of $P_{\textrm{indv.}}$ for the heterophilic empirical networks [Brazilian sexual contact network (Brazil), Swedish online dating network (POK), Facebook network of an university (USF51)] and the right two columns show the biases for the homophilic empirical networks [GitHub developers' network (GitHub), DBLP developers' network (DBLP), American Physical Society  network (APS)]. The gray crosses represent the perception bias of each individual estimated from the empirical network, and orange circles show the perception bias obtained from the BA-homophily model. The model reproduces well the heterogeneous distribution of individual perception bias in most of the networks. The simulation results are aggregated over $50$ iterations and the network size is $N=2,000$. The \textit{x} axis is shown in log scale. }
\label{fig:indi_wrd}
\end{figure*}

\clearpage

\subsection*{3. Growth rate ($\bm{C}$) in BA-homophily model}\label{subsec:C}
On the basis of the derivation provided by Karimi et al.~\cite{karimi2017visibility}, we can derive the exact degree dynamics of the BA-homophily model. Let us assume $K_a(t)$ and $K_b(t)$ as the sum of the degrees of each group $a$ and $b$, respectively. With the number of links a node has, the total number of links at each time step will be $K(t) = K_a(t) + K_b(t) = 2mt$. One can also describe the evolution of each group's degree as

\begin{align}
    \begin{dcases}
        K_a(t+\Delta t) = K_a(t) + m\Big( f_a \big( 1 + \frac{h_{aa}K_a(t)}{h_{aa}K_a(t)+h_{ab}K_b(t)} \big) + f_b \frac{h_{ba}K_a(t)}{h_{bb}K_b(t) + h_{ba}K_a(t)} \Big)\Delta t,\\
        K_b(t+\Delta t) = K_b(t) + m\Big( f_b \big( 1+ \frac{h_{bb}K_b(t)}{h_{bb}K_b(t) + h_{ba}K_a(t)} \big) + f_a \frac{h_{ab}K_b(t)}{h_{aa}K_a(t) + h_{ab}K_b(t)} \Big)\Delta t.\\
    \end{dcases}
\end{align}
\label{eq_sp:largeK}

Here, one can separate the amount of increase of the degree for each group with the limit $\Delta t \rightarrow 0$, 

\begin{align}
    \label{eq_sp:dk}
    \begin{dcases}
    \frac{dK_a}{dt} =  m\Big( f_a \big( 1 + \frac{h_{aa}K_a(t)}{h_{aa}K_a(t)+h_{ab}K_b(t)} \big) + f_b \frac{h_{ba}K_a(t)}{h_{bb}K_b(t) + h_{ba}K_a(t)} \Big),\\
    \frac{dK_b}{dt} = m\Big( f_b \big( 1+ \frac{h_{bb}K_b(t)}{h_{bb}K_b(t) + h_{ba}K_a(t)} \big) + f_a \frac{h_{ab}K_b(t)}{h_{aa}K_a(t) + h_{ab}K_b(t)} \Big).\\
    \end{dcases}
\end{align}

We can assume that $K_a(t)$ ($K_b(t)$) grows as a linear function of time. Given this assumption, we can write that $K_a(t) = Cmt$ ($K_b(t) = (2-C)mt$) based on $K(t) = 2mt$.

\begin{equation}
    \begin{dcases}
    \label{eq_sp:dka}
    \frac{dK_a}{dt} = Cm = m\Big( f_a \big( 1 + \frac{h_{aa}Cmt}{h_{aa}Cmt+h_{ab}(2-C)mt} \big) + f_b \frac{h_{ba}Cmt}{h_{bb}(2-C)mt + h_{ba}Cmt} \Big),\\
    \frac{dK_b}{dt} = (2-C)m = m\Big( f_b \big( 1+ \frac{h_{bb}(2-C)mt}{h_{bb}(2-C)mt + h_{ba}Cmt} \big) + f_a \frac{h_{ab}(2-C)mt}{h_{aa}Cmt + h_{ab}(2-C)mt} \Big).
    \end{dcases}
\end{equation}

Then, we can derive the relation of $C$ with group sizes and edge density in a group (Eq.~\ref{eq:C}) from Eq.~\ref{eq_sp:dk} by using $p_{aa}$ (Eq.~\ref{eq:paa}) and $p_{ab}$ (Eq.~\ref{eq:pab}).

\begin{figure}[ht]
\centering
\includegraphics[width=0.7\linewidth]{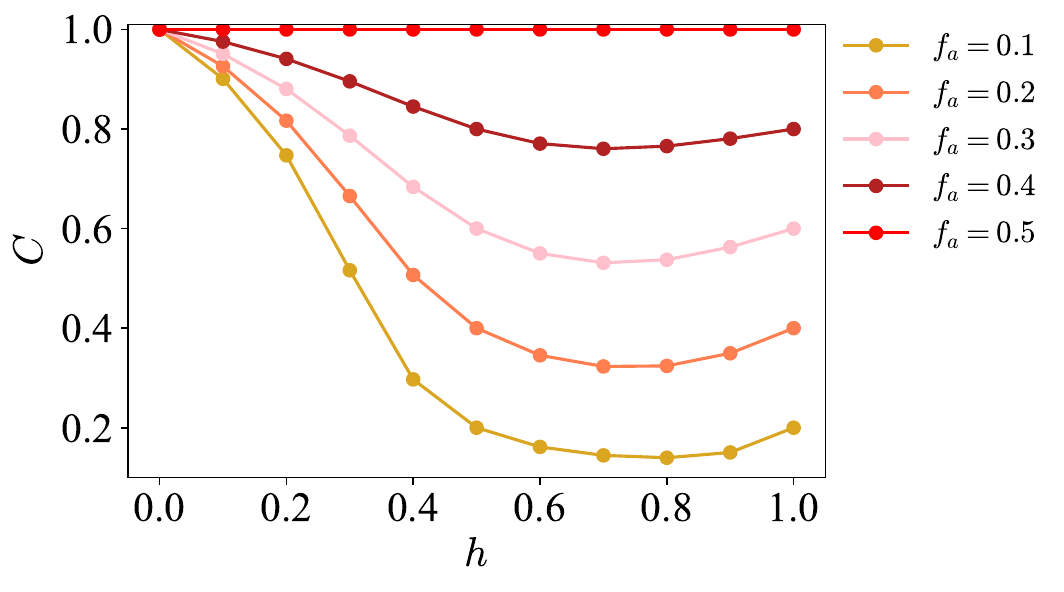} \\
\caption{\textbf{The relation between the minority's degree growth rate ($\bm{C}$) and the symmetric homophily ($\bm{h}$)}. As the minority group's size $f_a$ decreases, the degree growth rate of the minority drastically decreases with increasing symmetric homophily $h$ ($h_{aa}=h_{bb}$). $C$ is a function of $h$ and $f_a$ as described in Eq.~\ref{eq:C}.}
\end{figure}

\clearpage

\subsection*{4. Relationship between symmetric homophily ($\bm{h}$) and Newman's assortativity ($\bm{q}$)}
The symmetric homophily is equivalent to Newman's assortativity measure ($q$)~\cite{newman2003mixing}. The latter corresponds directly to the homophily parameter in the BA-homophily model after adjusting for scale in a relation $q = 2h - 1$. In the BA-homophily model, $h = 0$ means complete heterophily ($q = -1$), $h=0.5$ indicates no relationship between network structure and attributes ($q = 0$), and $h = 1$ indicates complete homophily ($q=1$). The relationship slightly deviates from the linear function for small minority-group sizes, but is independent of group sizes when $h \approx 0.5$. 

\begin{figure}[ht]
\centering
\includegraphics[width=0.5\linewidth]{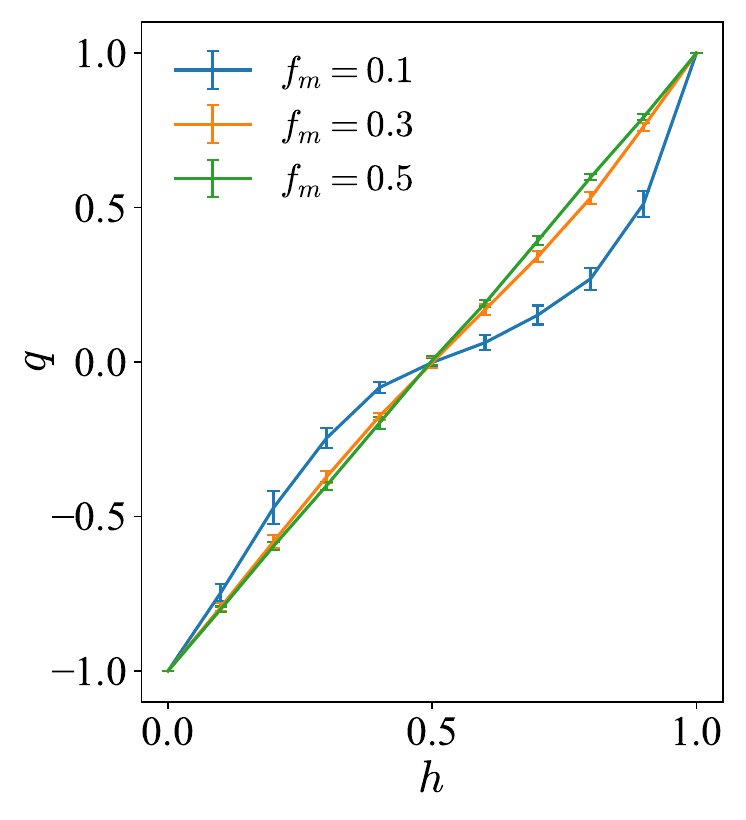} \\
\caption{\textbf{Relationship between Newman's assortativity ($\bm{q}$) and symmetric homophily ($\bm{h}$) in the BA-homophily model, for different sizes of minority group ($\bm{f_a}$).} Newman's assortativity is proportional to $h$ scaled as $2h-1$.  The relationship slightly deviates from the linear function for small minority-group sizes, but is independent of group sizes when $h \approx 0.5$.}
\label{fig:newman_assort}
\end{figure}
\clearpage

\subsection*{5. Bias of individual perceptions aggregated with those of 1-hop neighbors, assuming asymmetric homophily}

Here, we investigate to what extent and under what structural conditions individuals can reduce their perception bias for the prevalence of an attribute in a population by asking their friends about their perceptions and integrating those perceptions with their own when homophily is asymmetric. We build on DeGroot's weighted belief formalization by aggregating an individual's perception (\textit{ego}) with the averaged perceptions of the individual’s direct neighbors (\textit{1-hop}) \cite{degroot1974reaching} with an assumption of asymmetric homophily in the BA-homophily model. 

\begin{figure*}[!htp]
\centering

\includegraphics[width=0.52\linewidth]{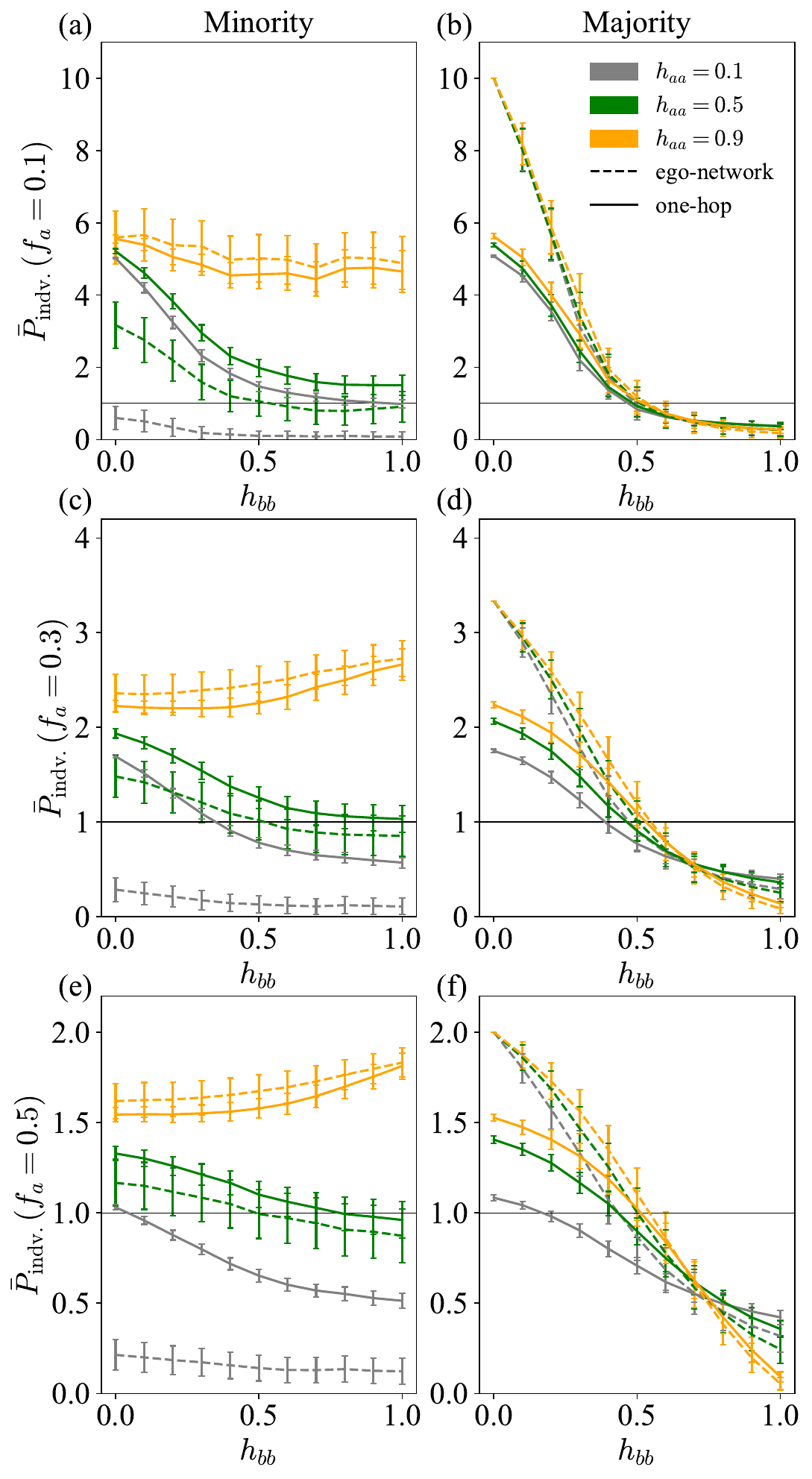}

\caption{\textbf{Individual-level perception biases (dashed lines) compared to the biases of the weighted average of perceptions of individuals and their 1-hop neighbors (solid lines), assuming asymmetric homophily}. The minority-group size is fixed at $0.1$ in the top row, $0.3$ in the middle row, and $0.5$ in the bottom row. The homophily for the minority $h_{aa}$ (depicted by lines of different colors) is fixed, while homophily for the majority $h_{bb}$ ranges from $0$ to $1$ along the horizontal axis.  
}
\label{fig:asym_onehop}
\end{figure*}

Figure \ref{fig:asym_onehop} shows a comparison of the average perception bias ($\bar{P}_{\textrm{indv.}}$) of individuals who are in (a) the minority and (b) the majority, with the bias of their perceptions aggregated with those of their 1-hop neighbors. The minority group size is fixed to $0.1$ in the top row, $0.3$ in the middle row, and $0.5$ in the bottom row. The homophily for the minority group $h_{aa}$ is fixed to $0.1, 0.5, 0.9$ (depicted by lines of different colors), while homophily for the majority group $h_{bb}$ ranges from $0$ to $1$ along the horizontal axis.

Results for asymmetric homophily (Fig.~\ref{fig:asym_onehop}) are generally similar to those for symmetric homophily (Fig.~\ref{fig:hop}): accounting for the opinion of 1-hop neighbors can decrease perception bias when networks are heterogeneous. For the majority, aggregating their own perceptions with those of their 1-hop neighbors leads to a robust improvement in perception accuracy, as described in Fig.~\ref{fig:asym_onehop}b, d, and f. For the minority, accounting for 1-hop neighbors also helps decrease the bias, though less than for the majority. However, when minority group is small and homophily is highly asymmetric ($h_{aa}=0.5$ and  $h_{bb}<0.5$), accounting for 1-hop neighbors can indeed increase the bias see Fig.~\ref{fig:asym_onehop}a).




\clearpage



\end{document}